\documentclass[11pt]{article}

\usepackage{epsfig,latexsym}

\newcommand{\sect}[1]{\setcounter{equation}{0}\section{#1}}

\def\aaa{angular momentum~}
\def\aaaa{angular momenta~}
\def\al {\alpha}

\def\AA{{\cal A}}

\def\ba{\begin{eqnarray}}

\def\bbbb{backgrounds~}
\def\be{\begin{equation}}
\def\bt {\beta}

\def\Bar {\overline}
\def\BB{{\cal B}}

\def\clll{conservation~laws~}

\def\CC{{\cal C}}

\def\de{\delta}
\def\di{\partial}

\def\De{\Delta}

\def\ea{\end{eqnarray}}
\def\ee{\end{equation}}

\def\emm{energy-momentum~}
\def\ep{\epsilon}

\def\fr{\frac}

\def\gmn{g_{\mu\nu}}
\def\grs{ g_{\rho\si}}
\def\sqg{\sqrt{-g}}
\def\Ga{\Gamma}

\def\ha{\frac{1}{2}}

\def\HH{{\cal H}}

\def\ka{\kappa}
\def\1k{\fr{1}{\ka}}
\def\2k{\fr{1}{2\ka}}
\def\4k{\fr{1}{4\ka}}

\def\la {\lambda}
\def\laa{\label}
\def\lll{\left(}

\def\La {\Lambda}

\def\LL{{\cal L}}
\def\LLL{\left[}

\def\MM{{\cal M}}

\def\na{\nabla}
\def\ni{\noindent}
\def\nl{\newline~}
\def\np{\newpage}

\def\OO {{\cal O}}

\def\pppp{perturbations~}

\def\ra{\rightarrow}
\def\rrr{\right)}

\def\Ra{\Rightarrow}

\def\RRR {\right]}

\def\ssss{spacetimes~}
\def\si{\sigma}
\def\suu{superpotential~}
\def\suuu{superpotentials~}

\def\Si{\Sigma}

\def\tbf{\textbf}
\def\td{\tilde}
\def\te{\theta}

\def\tsl{\textsl}

\def\TT{{\cal T}}

\def\UU{{\cal U}}

\def\vs{ \vskip 0.3 cm}
\def\ze{\zeta}

\title{{\bf Relativistic conservation laws and integral\\ constraints for large
cosmological perturbations}}
\author{ Joseph Katz\thanks{Permanent address: The Racah Institute of
Physics, 91904 Jerusalem, Israel. email: jkatz@phys.huji.ac.il},     Ji\v r\'\i \,
 Bi\v c\'ak\thanks{Permanent address: Institute of Theoretical Physics,
Faculty of Mathematics and Physics, Charles University, V Hole\v sovi\v
ck\'ach 2, 18000 Prague 8, Czech Republic. email:
bicak@mbox.troja.mff.cuni.cz},   and  Donald Lynden-Bell\thanks{email:
dlb@ast.cam.ac.uk} 
\\{\it  Institute of Astronomy, Madingley Road, Cambridge CB3 0HA, United
Kingdom}\\
\nl\\
Published in: Phys. Rev. D {\bf 55}, 5957 (1997)
\\(Received 19 July 1996)  }

\begin{document}

\maketitle
%_______________________ABSTRACT_____________________
\begin{abstract}

	For every mapping of a perturbed spacetime onto a background and with
any vector field $\xi$ we construct a   conserved
covariant vector density $I(\xi)$, which is the divergence of a covariant
antisymmetric tensor density, a ``superpotential". $ I(\xi)$  is linear in the
energy-momentum tensor perturbations of matter, which may be large;
$I(\xi)$ does not contain the second order derivatives of the perturbed
metric. The superpotential is identically zero when perturbations are absent. 
	By integrating conserved vectors over a part $\Si$ of a
hypersurface $S$ of the background, which spans a two-surface $\di\Si$,
we obtain integral relations between, on the one hand, initial data of the
perturbed metric components and the energy-momentum perturbations
on $\Si$  and, on the other hand, the boundary values on $\di\Si$. We
show that there are as many such integral relations as there are different
mappings,
$\xi$'s,
$\Si$'s and $\di\Si$'s. For given boundary  values on $\di\Si$, the integral
relations may be interpreted as integral constraints on local initial data
including the energy-momentum perturbations.
Conservation laws expressed in terms of Killing fields $\Bar\xi$ of
the background become ``physical" conservation laws. 
	In cosmology, to each mapping of the time axis of a Robertson-Walker space
on a de Sitter space with the same spatial topology there correspond ten
conservation laws. The conformal mapping leads to a straightforward
generalization of conservation laws in flat spacetimes. Other mappings are
also considered. 
	Traschen's ``integral constraints" for linearized spatially localized
perturbations of the energy-momentum tensor are examples of
conservation laws with peculiar $\xi$ vectors whose equations are rederived
here. In Robertson-Walker spacetimes, the ``integral constraint vectors" are
the Killing vectors of a de Sitter background for a special mapping.
[S0556-2821(97)00310-X]
\nl PACS number(s): 04.20.Cv, 98.80.Hw
\end{abstract}
 \np
%___________________I. INTRODUCTION __________________
\sect{Introduction}
 
%_______________A. STRONG CONSERVATION LAWS ETC... ETC... _______

\centerline{{\bf A. Strong conservation laws and cosmology}}
\vs 

Background spacetimes are commonly used in perturbation theories in
general relativity \cite{1} and play an essential role in cosmology \cite{2}.
One ``puzzle" \cite{3} in the theory of cosmological \pppp is Traschen's
``integral constraints" for {\it spatially localized} \pppp [4,5].
These Gauss-type restrictions on the \emm of matter \pppp
have significant effects \cite{6}: They point to an important reduction of the
Sachs-Wolfe \cite{7} effect on the mean square angular fluctuations at large
angles of the cosmic background temperature due to local inhomogeneities
in the universe for spatially isolated perturbations. 	

Traschen's relations remind us of Bergmann's {\it strong \clll} 
\cite{8} applied to perturbations of isolated systems. Such conservation
laws, which were explored in detail by Bergmann and Schiller \cite{9}, are, in
fact, identities. The identities, which involve an arbitrary vector $\xi$,
have played a basic role in the derivation of weak or N\oe ther conserved
currents in general relativity \cite{10} and are still in use \cite{11}. We
found it thus interesting to study   conservation laws on background
spacetimes \cite{12} in the context of cosmological perturbations. 

 Conservation laws are obtained from Lagrangians that are {\it scalar
densities} with not higher than first order derivatives of the fields. There are
{\it no} such {\it scalar
densities} for the metric and therefore  \clll in
general relativity are coordinate dependent. The coordinate dependence can
be ``brushed under the rug" by mapping the spacetime on a flat background
\cite{13}. This method offers, for example, the advantage of making the
Bondi mass \cite{14} calculable from Einstein's pseudotensor in Bondi
coordinates \cite{15} rather than in Minkowski coordinates \cite{16}. But
backgrounds are more than a useful tool in relativistic cosmology; they are
inevitable in linear and nonlinear perturbation theories.

Here, we derive strong \clll {\it with respect to curved \bbbb } along the
line indicated by Bergmann. We define a Lagrangian density  ${\hat L}_G$ 
for the gravitational field, quadratic in the first order covariant derivatives
 of the perturbed metric (the caret   means ``density", i.e., multiplication
by
$\sqrt{-g}$). $\hat L_G$   is normalized so that $\hat L_G=0$  when there
are no perturbations. Perturbations do not have to be small. The 
\clll derived from $\hat L_G$   are identically conserved vector
densities $\hat I^\mu(\xi)$,  the divergences of covariant superpotential
densities $\hat J^{\mu\nu}$ :
\be
\hat I^\mu =\di_\nu \hat J^{\mu\nu} ~~~   ,   ~~~ \hat J^{\mu\nu}  =-
\hat J^{\nu\mu}.
\label{11}
\ee
The $\hat I^\mu$ 's are identically conserved independently of whether
Einstein's equations are satisfied or not. However, we consider only metrics
that satisfy Einstein's equations.
$\hat I^\mu$'s are linear in the perturbed \emm tensor, and both $\hat
I^\mu$  and $\hat J^{\mu\nu}$  contain the perturbed metric and its
first-order covariant derivatives (no second-order derivatives); both are zero
when there are no perturbations. It follows from Eq. (\ref{11}) that if $\Si$ is
any piece of a hypersurface $S$ which spans a two-surface $\di \Si$,
\be
	\int_\Si\hat I^\mu d\Si_\mu=\int_{\di\Si}\hat I^{\mu\nu}
d\Si_{\mu\nu}.
\label{12}
\ee
These exact nonlinear integral identities represent  global conservation
laws if the integration is over the whole hypersurface $S$. If $\Si$ is only a
piece of the total, one may, in the manner of Penrose \cite{17}, speak of
quasi-local conservation laws. 
	
Now suppose that the boundary values, and thus $\hat J^{\mu\nu}$,  on
$\di\Si$ are {\it given}. Then Eq. (\ref{12}) represents an {\it integral
constraint} on the \pppp of the \emm tensor $\de T^\mu_\nu$  for
given initial perturbations of the metric on $\Si$. Reciprocally, if $\de
T^\mu_\nu$  is given, Eq. (\ref{12}) represents integral constraints on the
initial metric data on $\Si$. There are many integral constraints: for any
mapping, any $\xi$, and any $\Si$ with  the same boundary values and the
same $\xi$ and its first derivatives on $\di\Si$. Integral constraints may be
useful to relate boundary values of the metric to the matter sources on
$\Si$. 

Coming back to Traschen's integral constraints for linear
perturbations,  these represent particular forms of Eq.\,(\ref{12}) with a
class of ``integral constraint vectors"\,$\xi^\mu=V^\mu$ (not necessarily
Killing vectors), for which (\ref{12}) reduces to
\be
	\int_\Si\de T^\mu_\nu \hat V^\nu d\Si_\mu=0.
\label{13}
\ee
Boundary contributions are by definition vanishing. These equations are the
 integral constraints on $\de T^\mu_\nu$  that Traschen \cite{4} and
Traschen and Eardley \cite{6} considered for spatially localized
perturbations on a Robertson-Walker background. They found that
Eq. (\ref{13}) reduces considerably the Sachs-Wolfe \cite{7} effect of $\de
T^\mu_\nu$  on the angular fluctuations of the cosmic background
radiation. Different boundary values may have less stringent effects.
 
%_______________B. NOETHER CONSERVATION ETC... _______
\vs 
\centerline{{\bf B. N\oe ther conservation laws on curved \bbbb}}
 \vs
In special relativity \cite{18} like in general relativity
[19,20],  when the arbitrary vector $\xi^\mu$  is replaced by
a Killing vector of the background, $\bar\xi^\mu$, the  conservation
laws become physical conservation laws. N\oe ther conserved vectors $\hat
J^\mu$  have a physical content analogous to \emm and 
\aaa conserved currents in electromagnetism. However, contrary to
electromagnetism, conserved gravitational currents cannot be made
gauge independent, i.e., independent of the mapping.
	
N\oe ther conservation laws can be applied to asymptotically flat
spacetimes. This subject is not dealt with here in detail but it is noteworthy
that our superpotential $\hat J^{\mu\nu}$ gives properly the
``standard" expression for total energy, linear and angular momentum at
spatial infinity \cite{19} {\it and} at null infinity \cite{21} found in the
literature \cite{22}. The global conservation laws, {\it in their superpotential
forms}, relate local quantities to boundary values and, if applied globally,
give physical interpretations to ``asymptotic parameters" of solutions. They
are also useful in cosmology. 
	 
%_________C.  NOETHER CONSERVATION LAWS IN COSMOLOGY_______
\vs
\centerline{{\bf C. N\oe ther conservation laws in cosmology}}
 
\vs
In cosmology, there are six N\oe ther conservation laws for \pppp in
a  Robertson-Walker background, corresponding to the six Killing vectors.
There are four non-Noether conservation laws for each of the remaining
conformal Killing vectors. The ten vectors correspond to the fact that
Robertson-Walker spacetimes are conformal to de Sitter spacetimes which, as
is well known, admit ten independent Killing vectors like Minkowski space.
In cosmological applications, de Sitter spaces appear more suitable as
backgrounds than Minkowski spaces. The more so because in inflationary
scenario, de Sitter spacetimes transform into Robertson-Walker spacetimes.
The quasi-energy and initial position of the mass center \cite{40} can be
associated with the four Killing vectors of de Sitter spaces which do not 
correspond to the six Killing vectors of Robertson-Walker universes. 

	 Traschen's integral constraints, which we mentioned before, look like
conservation laws on a de Sitter space in disguise. Todd \cite{23} has shown
that the equations for {\it integral constraint vectors} (or ICV's) $V^\mu$
are conditions for $\Si$ to be embeddable in a spacetime with constant
curvature of which the  $V^\mu$'s are the Killing vectors.

In Section II we give the general theory of strong conservation laws relative
to a curved background for both nonlinear and linearized perturbations.
A summary of some of the results appeared already in \cite{12}.  Here we
give full details and we also include a generalization of the
Belinfante-Rosenfeld identities \cite{24}. Section III is devoted to Noether
conservation laws; the energy-momentum tensor and helicity tensor with
respect to the background are singled out. Results of applications to
asymptotically flat backgrounds are mentioned. Section IV gives details on
Noether's conservation laws for Robertson-Walker spaces mapped on de
Sitter spaces with the same spatial topology. In Section V Traschen's integral
constraints are related to  conservation laws. Integral constraint
vectors are shown to be the Killing vectors of a de Sitter background with a
particular mapping.
 
%___________________II. STRONGLY CONSERVED CURRENTS__________

\sect {Strongly conserved currents}

The main result of this section is summarized in Eq.\,(\ref{239}).
\vs 
\centerline{	\bf{A.  Lagrangian density for gravitational fields on a curved
background}}
 \vs
	Let $g_{\mu\nu}(x^\la) , ~\la, \mu,\nu, ...= 0,1,2,3,$ be the metric of a
spacetime $\MM$  with signature $-2$, and let 
$\Bar g_{\mu\nu}(\Bar x^\la)$  be the metric of the background
$\Bar\MM$. Both are tensors with respect to arbitrary coordinate
transformations. Once we have chosen a mapping so that points $P$ of
$\MM$ map into points $\Bar P$ of  $\Bar\MM$ , then we can use the
convention that $\Bar P$ and $P$ shall always be given the same coordinates
$\Bar x^\la =x^\la$. This convention implies that a coordinate
transformation on
$\MM$ inevitably induces a coordinate transformation with the same
functions on $\Bar\MM$. With this convention, such expressions as
$\gmn(x^\la)-\Bar\gmn(x^\la)$  become true tensors. However, if
the particular mapping has been left unspecified we are still free to change
it. The form of the equations for perturbations must inevitably contain a
gauge invariance corresponding to this freedom.

 Let $R^\la_{~\nu\rho\si}$,
and $\Bar R^\la_{~\nu\rho\si}$  be the curvature tensors of $\MM$ and
$\Bar\MM$. These are related as follows \cite{25}:
\be
	R^\la_{~\nu\rho\si} = \Bar D_\rho \De ^\la_{\nu\si}- \Bar
D_\si\De^\la_{\nu\rho} + \De^\la _{\rho\eta}\De^\eta_{\nu\si} -
\De^\la _{\si\eta}\De^\eta_{\nu\rho} + \Bar R^\la_{~\nu\rho\si}.
\label{21}
\ee
Here $\Bar D_\rho$  are covariant derivatives with respect to $\Bar\gmn$ 
and
$\De^\la _{\mu\nu}$  is the difference between Christoffel symbols in
$\MM$  and $\Bar\MM$ :
\be
\De^\la _{\mu\nu}=\Ga^\la _{\mu\nu}-\Bar\Ga^\la _{\mu\nu}=\ha
g^{\la\rho}\lll  \Bar D_\mu g_{\rho\nu}+ \Bar D_\nu g_{\rho\mu}-
 \Bar D_\rho \gmn\rrr. 
\label{22}
\ee
 Our quadratic Lagrangian density  $\hat\LL_G $   for gravitational
perturbations is then defined as 
\be
\hat\LL_G  =\hat\LL-\Bar{\hat\LL }~~~,~~~  \hat\LL = -\fr{1}{2\ka}(\hat
R+\di_\mu\hat k^\mu) ~~~ , ~~~ \Bar{\hat\LL} = -\fr{1}{2\ka}\Bar{\hat
R}~~~  ,  ~~~
\ka =\fr{8\pi G}{c^4}.
\label{23} 
\ee
The caret  $\hat {}$ means, as we said before, multiplication by  $\sqrt{-g}$ ,
never by $\sqrt{-\Bar {g}}$.   Thus, if $\hat{R}  = \sqrt{-g} R $, $\Bar{\hat
R}$ will unambiguously mean$\sqrt{-\Bar g} \Bar R $ . Notice that 
$\hat{\Bar R}=\sqrt{-g}\Bar R\neq \Bar{\hat R}=\sqrt{-\Bar g} \hat{\Bar R
}/\sqrt{-g}$. The divergence of the vector density
$\hat k^\mu$,
\be
	\hat k^\mu  = \fr{1}{\sqrt{-g}}\Bar D_\nu(-gg^{\mu\nu})=\hat
g^{\mu\rho}\De^\si_{\rho\si}- \hat g^{ \rho\si}\De^\mu_{\rho\si},
\label{24}		
\ee
cancels all second order derivatives of $\gmn$  in $R$. $\hat\LL$  is the
Lagrangian used by Rosen. $\Bar{\hat{\LL}}$  is $\hat\LL$  in which
$\gmn$  has been replaced by $\Bar\gmn$. When $\gmn=\Bar\gmn$,
$\hat\LL_G$  is thus identically zero. The intention here is to obtain
\clll  in the background space so that if $\gmn=\Bar\gmn$, conserved
vectors and superpotentials would be identically zero as in Minkowski space
in special relativity. The following formula, deduced from Eq.  (\ref{23}) and
Eq. (\ref{21}), shows explicitly how  $\hat\LL_G$  is quadratic in the first
order derivatives of $\gmn$  or, equivalently, quadratic in
$\De^\mu_{\rho\si} $:
\be
	\hat\LL_G  = \fr{1}{2\ka}\hat g^{\mu\nu}\lll  \De^\rho_{\mu\nu} 
\De^\si_{\rho\si} - \De^\rho_{\mu\si}  \De^\si_{\rho\nu} \rrr - 
\fr{1}{2\ka}(\hat g^{\mu\nu}-\Bar{\hat g^{\mu\nu}})\Bar R_{\mu\nu}.	 
\label{25}
\ee
Notice that if $\Bar R^\la_{~\nu\rho\si}=0$ and coordinates are such that    
$\Bar \Ga^\la_{\mu\nu}=0$, $\hat\LL_G$ is nothing else than the familiar
``$\Ga\Ga-\Ga\Ga$" Lagrangian density \cite{26}.

%________________B. INFINITESIMAL REPARAMETRIZATION ETC...______

 \vs
\centerline{	\bf{B. Infinitesimal reparametrization in both $\MM$ and
$\Bar\MM$ }}
 \vs
Lie differentials are particularly convenient in describing infinitesimal
displacements in both $\MM$ and $\Bar\MM$; they are thus not associated
with a change of mapping. If	
\be
		\De x^\mu =\xi^\mu \De\la	
\label{26}
\ee
represents an infinitesimal one-parameter displacement generated by  a
sufficiently smooth vector field $\xi^\mu$, the corresponding changes in
tensors are given in terms of the Lie derivatives with respect to the vector
field $\xi^\mu$, $\De\gmn = \pounds  
_\xi\gmn
\De\la$, etc\footnote{Here the symbol $``\De\la"$
denotes an infinitesimal quantity. It has no direct connection with 
$\De^\rho_{\mu\nu}$ above which is finite.}. The Lie derivatives may be written in terms of ordinary partial
derivatives
$\di_\mu$, covariant derivative $\Bar D_\mu$  with respect to
$\Bar\gmn$, or covariant derivative $D_\mu$ with respect to $\gmn$.
Thus,
\ba
\pounds_\xi \gmn& =& g_{\mu\la}\di_\nu\xi^\la +g_{\nu\la}\di_\mu\xi^\la
+\xi^\la\di_\la g_{\mu\nu}\nonumber\\
& =&g_{\mu\la}\Bar D_\nu\xi^\la +g_{\nu\la}
\Bar D_\mu\xi^\la+\xi^\la \Bar D_\la g_{\mu\nu}= g_{\mu\la}D_\nu\xi^\la
+g_{\nu\la}D_\mu\xi^\la.
\label{27}
\ea
	
Consider now the Lie differential $\De\hat\LL$  of $\hat\LL$  . With the
variational principle in mind, we write $\De\hat\LL=\pounds_\xi \hat\LL\De\la$ in
the form
\be
\De\hat\LL=\fr{1}{2\ka}\hat G^{\mu\nu}\De\gmn+\di_\mu\hat
A^\mu\De\la,
\label{28}
\ee
where Einstein's tensor density, $\hat G^{\mu\nu}=\hat R^{\mu\nu}-\ha
 \hat R  g^{\mu\nu}$, is the variational derivative of $ \hat\LL$  with
respect to $\gmn$ and $\hat A^\mu$  is a vector density linear in ${\bf
\xi}$ (see below). The Lie derivative of a scalar density like $ \hat\LL$  is just
an ordinary divergence $\di_\mu( \hat\LL\xi^\mu)$. Thus 
\be
	\hat	\OO \equiv   \pounds_\xi \hat\LL  - \di_\mu( \hat\LL\xi^\mu)=0.	 
\label{29}
\ee
Combining Eq. (\ref{29}) with Eq. (\ref{28}), we obtain
\be
\hat	\OO \equiv  \fr{1}{2\ka}\hat G^{\mu\nu} \pounds_\xi \gmn + \di_\mu\hat
B^\mu=0, 
\label{210}
\ee
 where
\be
\hat B^\mu= \hat A^\mu  -\hat\LL\xi^\mu  = \ha\hat
\Si^{\mu\rho\si}\pounds_\xi g_{\rho\si} + \hat \Xi ^\mu 
-\hat\LL\xi^\mu,
\label{211}
\ee
  		 in which
\be
2\ka \hat \Si^{\mu\rho\si}= \lll g^{\mu\rho} g^{\si\nu} +g^{\mu\si}
g^{\rho\nu}  - g^{\mu\nu} g^{ \rho\si}
\rrr \hat\De^\la_{\nu\la}  -\lll g^{\nu\rho} g^{\si\la} +g^{\nu\si}
g^{ \rho\la}  -g^{\nu\la} g^{\rho\si} \rrr\hat\De^\mu_{\nu\la}
\label{212}
\ee
and
\be
	4\ka \hat \Xi ^\mu  = \hat g^{\mu\la} \di_\la Z+\hat g^{\rho\si} \LLL
\Bar D^\mu Z_{\rho\si} -\lll \Bar D_\rho Z^\mu_\si +\Bar D_\si
Z^\mu_\rho\rrr\RRR,
\label{213}
\ee
 with
\be
Z_{\rho\si}\equiv \pounds_\xi\Bar g_{\rho\si}=\Bar D_\rho\xi_\si+\Bar D
_\si\xi_\rho~~~,~~~Z=\Bar g^{\rho\si}Z_{\rho\si}~~~,~~~\xi_\si=\Bar g_{\si\mu}\xi^\mu.
\label{214}
\ee	
Hereafter, indices are moved up or down with $\Bar \grs$  only, never
with
$\grs$. In Eq. (\ref{211}), $\pounds_\xi \grs$  may be replaced by its
expression Eq. (\ref{27}) in terms of $\Bar D_\nu$ derivatives. In this way,
$\hat B^\mu$  contains $\Bar D_\nu$ derivatives only.

	Belinfante \cite{24} and Rosenfeld \cite{24} extracted from Eq. (\ref{210})
various identities and showed how to complete Pauli's canonical \emm
tensor to make it symmetrical \cite{27}. Identities   (\ref{210}) have been
used to construct strong conservation laws in general relativity, without
mapping on a background \cite{28} and, more rarely, with a mapping on a
flat background \cite{19}. Here we use the identities Eq. (\ref{210}) to
construct strong \clll on curved backgrounds. Bianchi identities imply
$D_\nu  G^{\mu\nu} =0~$ so that with Eq. (\ref{27}), Eq. (\ref{210}) can be
written as the divergence of a vector density 
\be
\hat\OO = \di_\mu \hat j^\mu =0~~~{ \rm    where }~~~  \hat j^\mu =  
\fr{1}{\ka}\hat G^\mu_\nu\xi^\nu+\hat B^\mu. 
\label{215}
\ee
Hence, $\hat \LL$  ``generates" a vector density $\hat j^\mu$  that is
{\it identically} conserved. It has been obtained without using Einstein's
field equations; Eq. (\ref{215}) is the kind of  ``strong"  conservation law
introduced by Bergmann \cite{8}. We shall, of course, assume that Einstein's
equations are satisfied, and replace  $(1/\ka)G^\mu_\nu$  by the
\emm of matter 
\be
	\1k G^\mu_\nu=T^\mu_\nu,
\label{216}
\ee
so that our conservation law  Eq. (\ref{215}) reads
\be
	\di_\mu\hat  j^\mu  =\di_\mu(\hat T^\mu_\nu \xi^\nu + \hat B^\mu)  =
0.	
\label{217}
\ee
Equation (\ref{217}) is, strictly speaking, not an identity anymore. Given
$\hat T^\mu_\nu$, Eq.  (\ref{217}) holds only for metrics that satisfy
Eq. (\ref{216}).
$\hat j^\mu$  is linear in $\xi$ and its derivatives up to order 2. If in $\hat
B^\mu$, the $\Bar D_\rho\xi_\si$  are decomposed into symmetric and
antisymmetric parts, using $Z_{\rho\si}$  defined in Eq. (\ref{214}),
\be
		\Bar D_\rho\xi_\si =\ha \lll  \Bar D_\rho\xi_\si- \Bar D_\si\xi_\rho
\rrr+ \ha \lll  \Bar D_\rho\xi_\si+ \Bar D_\si\xi_\rho
\rrr\equiv\di_{[\rho}\xi_{\si]}  + 
\ha Z _{\rho\si},	
\laa{218}
\ee
$\hat j^\mu$  takes the following form:
\be
	\hat j^\mu = \hat P^\mu_\nu  \xi^\nu  + \hat\si ^{\mu [\rho\si]}
\di_{[\rho}\xi_{\si]}   + \hat Z^\mu,
\laa{219}
\ee
 in which
\be
	\hat P^\mu_\nu =\hat T^\mu_\nu +  \fr{1}{2\ka}\hat g^{\rho\si}\Bar R_{\rho\si}\de^\mu_\nu +\hat t^\mu_\nu ,	
\laa{220}
\ee
 with 
\ba
	2\ka \hat t^\mu_\nu  &= & \hat g^{\rho\si} \LLL (\De^\la_{\rho\la}
\De^\mu_{\si\nu} +\De^\mu_{\rho\si}\De^\la_{\la\nu}
-2\De^\mu_{\rho\la}\De^\la_{\si\nu}
)-\de^\mu_\nu (\De^\eta_{\rho\si}
\De^\la_{\eta\la}-\De^\eta_{\rho\la}\De^\la_{\eta\si})\RRR
\nonumber\\&+&
 \hat g^{\mu\la}(\De^\si_{\rho\si}\De^\rho_{\la\nu}-\De^\si_{\la\si}
\De^\rho_{\rho\nu}),	
\laa{221}
\ea
and $\hat\si ^{\mu [\rho\si]}$  is the antisymmetric part of $\hat\si ^{\mu
\rho\si}$  related to $\hat\Si ^{\mu\rho\si}$  as follows 
\ba
2\ka\hat\si ^{\mu
\rho\si}&  = &2\ka\hat\Si ^{\mu\rho\la}  g_{\la\nu} \Bar g^{\nu\si } = \lll
g^{\mu\rho} \Bar g^{\si\nu}+\Bar g^{\mu\si}   g^{\rho\nu} 
- g^{\mu\nu}\Bar g^{\rho\si }\rrr\hat 
\De^\la_{\nu\la}\nonumber\\  
			                      & -& \lll g^{\nu\rho} \Bar g^{\si\la} +\Bar
g^{\nu\si}g^{\rho\la}   -g^{\nu\la} \Bar g^{\rho\si}\rrr
\hat \De^\mu_{\nu\la}	 
\laa{222}
\ea
 (the terms containing $ \Bar g^{\rho\si}$  do not contribute to
$\hat\si ^{\mu [\rho\si]}$) while 
\ba
	4\ka \hat Z^\mu & =& (Z^\mu_\rho \hat g^{\rho\si}  +\hat
g^{\mu\rho }Z^\si_\rho-\hat g^{\mu\si}Z )\De^\la_{\si\la} +(\hat
g^{\rho\si}Z -2\hat g^{\rho\la}Z^\si_\la )\De ^\mu_{\rho\si} \nonumber\\
			                                  & +&\hat g^{\mu\la}\di_\la
Z+\hat g^{\rho\si}(\Bar D^\mu Z_{\rho\si}-2\Bar D_\rho Z^\mu_\si) .	
\laa{223}
\ea
 
%________________C. SUPERPOTENTIALS AND ETC...______
\vs
 
\centerline{	\bf{C. Superpotentials and strong conservation laws}}
\vs

	Since $ \hat j^\mu$  as given by  Eq. (\ref{215}) is identically conserved
whatever  $\gmn$ is, it must be the divergence of an antisymmetric tensor
density that depends on arbitrary $\gmn$'s too;  thus,
\be
	 \hat j^\mu= \di_\nu  \hat j^{\mu\nu } , ~~~   {\rm where }~~~   \hat
j^{\mu\nu }= -\hat j^{\nu\mu },
\label{224}
\ee
$\hat j^{ \mu\nu }$  is easy to find and has been derived directly
from $\hat\LL$  in \cite{19}. In those papers the background is assumed to
be flat, but the derivation of
$\hat j^{\mu\nu}$ does not depend on that assumption: 
\be
		j^{\mu\nu}  = \fr{1}{\ka} D^{[\mu} \xi^{\nu]} + \fr{1}{\ka}
\xi^{[\mu}k^{\nu]}.	
\label{225}
\ee
The terms  $\fr{1}{\ka} D^{[\mu} \xi^{\nu]}$ will be recognized as 
$\ha$ Komar's
superpotentials \cite{29}. In terms of $\Bar D$  derivatives, 
\be
		D_\rho\xi^\mu  = \Bar D_\rho\xi^\mu  + \De^\mu_{\rho\la} \xi^\la,		
\label{226}
\ee
and regarding expression  (\ref{24}) for $k^\mu ,  j^{\mu\nu}$  may be
written in the form
\be
	\ka j^{\mu\nu} = g^{\rho[\mu} \Bar D_\rho \xi^{\nu]} + g^{\rho[\mu} 
\De^{\nu]}_{\rho\la}  \xi^\la  + \xi^{[\mu} g^{\nu]\rho} \De
^\si_{\rho\si}   - \xi^{[\mu} \De^{\nu]}_{\rho\si} g^{\rho\si} .
\label{227}
\ee
	
Had we applied the identities Eq. (\ref{29}) to $\Bar{\hat\LL}$  instead of
$\hat\LL$, we would have written everywhere $\Bar\gmn$  instead of
$\gmn$, from Eq. (\ref{29}) up to Eq. (\ref{223}). We would have found
strong, barred, conserved vector densities $\Bar{\hat j^\mu}$  and barred
\suuu $\Bar{\hat j^{\mu\nu}}$:	
\be
\Bar{\hat j^\mu} = \lll \Bar{\hat T^\mu _\nu} +\fr{1}{2\ka} \Bar{\hat
R}\de^\mu_\nu \rrr \xi^\nu+\Bar{\hat Z^\mu} = \di_\nu\Bar{\hat
j^{\mu\nu}},
\label{228}
\ee
with
\be
\Bar{\hat Z^\mu}  = \Bar{\hat g^{\mu\la}} \di_\la Z+\Bar{
\hat g^{\rho\si}}(\Bar D^\mu Z_{\rho\si}-2\Bar D_\rho Z^\mu_\si),	
\label{229}
\ee
and
\be  
\Bar{\hat
j^{\mu\nu}} = \fr{1}{\ka}  \Bar{D^{[\mu}\hat\xi^{\nu]}}.
\label{230}
\ee
Strongly conserved vectors for $\hat\LL_G  =\hat\LL-\Bar{\hat\LL }$   are
thus obtained by subtracting barred vectors and superpotentials from
unbarred ones; in this way, we define relative vectors and in particular
{\it relative \suuu} $\hat J^{\mu\nu}$-- relative to the background space.
Setting
\be
\hat I^\mu =\hat j^\mu-\Bar {\hat j^\mu }~~~      ,   ~~~    \hat 
J^{\mu\nu} =\hat j^{\mu\nu}-\Bar {\hat j^{\mu\nu}} = -\hat 
J^{\nu\mu},	
\label{231}
\ee
we have
\be 
		 \hat I^\mu\equiv \hat J^\mu  + \hat\ze^\mu  =\di_\nu \hat
J^{\mu\nu}   ~~~,~~~ \di_\mu \hat I^\mu\equiv0,		
\label{232}
\ee
 where 
\be
		      \hat J^\mu = \hat\te^\mu_\nu \xi^\nu+\hat
\si^{\mu[\rho\si]}\di_{[\rho}\xi_{\si]},
\label{233}
\ee
with 
\be
 		      \hat \te^\mu_\nu= \de \hat T^\mu_\nu  + \fr{1}{2\ka} \hat
l^{\rho\si}  \Bar R_{\rho\si}\de^\mu_\nu + \hat t^\mu_\nu,		
\label{234}
\ee
 in which 
\be
	\de \hat T^\mu_\nu \equiv  \hat T^\mu_\nu   - \Bar{\hat
T^\mu_\nu}~~~,~~~ \hat l^{\mu\nu} \equiv  \hat g^{\mu\nu}   -
\Bar{\hat g^{\mu\nu}},	
\label{235}
\ee
and  $\hat\ze^\mu = \hat Z^\mu  -\Bar{\hat Z^\mu}$   is given by
\ba
4\ka \hat\ze^\mu &=& \lll  Z^\mu_\rho  \hat  g^{\rho\si}
+ \hat g^{\mu\rho}
 Z_\rho^\si -  \hat g^{\mu\si} Z\rrr \De^\la_{ \si\la} +\lll  \hat  g^{\rho\si}
 Z-2 \hat g^{\rho\la}  Z_\la^\si
\rrr \De^\mu_{\rho\si} \nonumber\\
&+& \hat l^{\mu\la}\di_\la   Z
	   + \hat l^{\rho\si}\lll \Bar D^\mu  Z_{\rho\si}-2\Bar D_\rho
  Z^\mu_\si\rrr,	
\label{236}
\ea
 while the
superpotential is given by 
\be
		\hat J^{\mu\nu} =  \fr{1}{\ka}\lll D^{[\mu}\hat \xi^{\nu]} -
\Bar{D^{[\mu}\hat \xi^{\nu]}} + \hat \xi^{[\mu} k^{\nu]}\rrr.		
\label{237}
\ee
$\hat J^{\mu\nu}$ can also be written in terms of $\gmn$ ,
$\De^\mu_{\rho\si}$  and $\xi^\mu$:
\be
	\ka\hat J^{\mu\nu}  = \hat l^{\rho[\mu} \Bar D_\rho\xi^{\nu]}+\hat
g^{\rho[\mu} \De^{\nu]}_{\rho\la}\xi^\la  + \xi^{[\mu}  \hat g^{\nu]\rho}
\De^\si_{\rho\si}  - \xi^{[\mu}\De^{ \nu]}_{\rho\si} \hat g^{\rho\si}.
\label{238}
\ee

	The tensors in Eq. (\ref{233}) have a physical interpretation. On a flat
background, in coordinates in which $\bar\Ga^\la_{\mu\nu} =0$, $\hat
t^\mu_\nu$ reduces to Einstein's pseudo-tensor. $\hat\te^\mu_\nu$ 
appears therefore as the \emm tensor of the \pppp
with respect to the background. The second tensor in Eq. (\ref{233}),
$\hat\si ^{\mu [\rho\si]}$, is quadratic in the metric perturbations just like
$\hat t^\mu_\nu$. It is also bilinear in the perturbed metric components
$(\gmn-\bar\gmn)$ and their first order derivatives. $\hat\si ^{\mu
[\rho\si]}$ resembles, in this respect, the helicity tensor density in 
electromagnetism (see below). The factor of $\di_{[\rho}\xi_{\si]}$
represents thus the helicity tensor density of the perturbations with respect
to the background.
	
It should be noted again that all the components of $I^\mu$  and of the
\suu $J^{\mu\nu}$  itself are identically zero if $\gmn=\Bar\gmn$;
therefore,   \clll refer to perturbations only and not to
the background.

	To summarize, the main result obtained so far is the explicit form of strongly
conserved vectors $\hat  I^\mu$  and their associated superpotentials $
\hat J^{\mu\nu}$ on any background :
\be
 \hat I^\mu \equiv \hat\te^\mu_\nu \xi^\nu  + \hat\si ^{\mu [\rho\si]}
\di_{[\rho}\xi_{\si]} + \hat \ze^\mu  = \di_\nu\hat J^{\mu\nu},		
\label{239}
\ee
in which $\hat\te^\mu_\nu$  is given in Eq. (\ref{234}), $\hat\si ^{\mu
[\rho\si]}$ in Eq. (\ref{222}), $\hat \ze^\mu $  in Eq. (\ref{236}), and $\hat
J^{\mu\nu}$  in Eq. (\ref{238}). $\hat I^\mu$  is strongly conserved for any
$\xi^\mu$  and any mapping of $\MM$ on $\Bar\MM$.
 
%________________D. STRONG INTEGRAL CONSERVATION ETC....______
\vs
 
\centerline{	\bf{D.  Integral conservation laws and integral constraints
}}
 \vs
We can now integrate Eq. (\ref{239}) on a part $\Si$ of a hypersurface $S$
which spans a two-surface $\di \Si$ and obtain a {\it  integral
conservation law}:
\be
	\int_\Si	 \lll    \hat\te^\mu_\nu \xi^\nu  + \hat\si ^{\mu [\rho\si]}
\di_{[\rho}\xi_{\si]} + \hat\ze^\mu\rrr d\Si_\mu  = 	\int_\Si\hat
J^{\mu\nu}d\Si_{\mu\nu}	.	
\label{240}
\ee
	On both sides of this equality appear, besides $\de \hat T^\mu_\nu$,
components of the metric perturbations and their first order derivatives.
Therefore, Eq. (\ref{240}) is an integral relation between possible metric
initial data on $\Si$, the energy-momentum perturbations $\de \hat
T^\mu_\nu$  and the boundary values on $\di\Si$. For {\it fixed} boundary
values, and for each $\xi^\mu$, Eq. (\ref{240}) gives an {\it integral
constraint} on the metric initial data for given $\de \hat
T^\mu_\nu$. Reciprocally, for given metric initial
data, Eq. (\ref{240}) is an integral constraint on $\de \hat
T^\mu_\nu$. In particular, if
\pppp are ``localized" in the sense that the boundary integral is
zero, then the integral constraints are simply given by 
\be
\int_\Si	 \lll    \hat\te^\mu_\nu \xi^\nu  + \hat\si ^{\mu [\rho\si]}
\di_{[\rho}\xi_{\si]} + \hat\ze^\mu\rrr d\Si_\mu  = 0 ~~~~{\rm      
(isolated~ system)}.	
\label{241}
\ee
	There exist special vectors $\xi^\mu$  for which the expression  
(\ref{236}) for $\ze^\mu$  takes a somewhat simpler form: 

 If the background admits
conformal Killing vectors, like in Robertson-Walker spacetimes,
\be
		Z_{\rho\si}  = \fr{1}{4} \Bar g_{\rho\si} Z,		
\label{242}
\ee
and Eq. (\ref{236}) becomes
\be
	8\ka \hat \ze^\mu =\lll \hat l^{\mu\rho}  + \ha \Bar g^{\mu\rho}
\hat l\rrr \di_\rho Z - \lll  \hat g^{\mu\rho} \De^\si_{\rho\si}-
\hat g^{\rho\si}\De^\mu_{\rho\si}\rrr Z ~~~    {\rm (\xi^\mu ~
conformal)}.	
\label{243}
\ee

 If $\xi^\mu$  is a homothetic Killing vector,
\be
		Z_{\rho\si}  =\fr{1}{4}\Bar g_{\rho\si} C ~~~   , ~~~    C= {\rm const}  ,
\label{244}
\ee
Eq. (\ref{243}) reduces to
\be
		8\ka\hat  \ze^\mu = -\lll \hat g^{\mu\rho}\De^\si_{\rho\si}
-\hat g^{\rho\si}
\De^\mu_{\rho\si} \rrr C = - C\hat k^\mu ~~~ {\rm (\xi^\mu~
homothetic)}.	
\label{245}
\ee

For Killing vectors of the background, which hereafter will be denoted by
$\Bar\xi^\mu$   we get $\ze^\mu=0$. If, in addition, Killing vectors are
tangent to $\Si$,  $\Bar \xi^\mu d\Si_\mu=0$, as will be the case in
Robertson-Walker spacetimes mapped on de Sitter spaces, the coupling to
the background Ricci tensor in
Eq. (\ref{234}) disappears, and Eq. (\ref{240}) reduces to
\be
 \int_\Si\LLL  \lll  \de\hat T^\mu_\nu+ \hat t^\mu_\nu \rrr \Bar\xi^\nu
+
\hat \si^{\mu[\rho\si]}\di_{[\rho}\Bar \xi_{\si]}
\RRR d\Si_\mu = \int_{\di\Si} \hat J^{\mu\nu}d\Si_{\mu\nu}~~~ ~~~   
(\Bar\xi^\mu d\Si_\mu =0). 
\label{246}
\ee

%________________E. BELINFANTE-ROSENFELD IDENTITIES.______
\vs
 
\centerline{	\bf{E. Belinfante-Rosenfeld identities}}
 \vs
Equation   (\ref{232}), $\di_\mu\hat  I^\mu =0$, with $\hat I^\mu$ 
depending linearly on $\xi^\mu$'s and their first order derivatives, holds
for any
$\xi^\mu$. Therefore, $\di_\mu \hat I^\mu =0$ is a linear combination of
the
$\xi^\mu$ 's and their derivatives $\Bar D_\la\xi^\mu$  and
$\Bar D_{(\rho\si)}\xi^\mu$:
\be
		\di_\mu\hat  I^\mu  = \hat\OO_\nu\xi^\nu +\hat\OO^\mu_\nu
\Bar D_\mu \xi^\nu + \hat\OO^{\rho\si}_\nu \Bar D_{(\rho\si)}
\xi^\nu\equiv 0,
\label{247}
\ee
whose coefficients must be identically zero. This gives 60 identities -- the
Belinfante-Rosenfeld identities generalized to curved backgrounds. 
Integral \clll and integral constraints are obtained with linear combinations
of these 60 identities with $\xi^\mu$ and its derivatives as coefficients. 
Calculations of the coefficients are somewhat tedious, but straightforward. A
useful equation is that which transforms $\ze^\mu$  into an expression
depending on $\Bar D_\mu \xi^\nu$  and $\Bar D_{(\rho\si)} \xi^\nu$
rather than $Z_{\rho\si}$  and $\Bar D_\la Z_{\rho\si}$:
\be
\hat\ze^\mu  = \lll - \fr{1}{4\ka} \hat l^{\mu\rho}\Bar R_{\rho\nu} +  
\fr{1}{2\ka} \hat l^{\rho\si}\Bar R^\mu_{~\rho\si\nu} \rrr\xi^\nu +
\hat\si^{\mu(\rho\si)}\Bar
D_{(\rho}\xi_{\si)}+\hat\bt_\la^{~\mu \rho\si}\Bar D_{(\rho\si)}\xi^\la,
\label{248} 
\ee
where $\hat\si^{\mu\rho\si}$  is defined in Eq. (\ref{222}), while
\be
	\hat\bt_\la^{~\mu\rho\si}=\fr{1}{4\ka}\lll  \hat l^{\mu\rho}
\de^\si_\la+\hat l^{\mu\si}
\de^\rho_\la-2\hat l^{\rho\si}
\de^\mu_\la\rrr=	\hat\bt_\la^{~\mu\si\rho}.	
\label{249}
\ee
	Inserting Eq. (\ref{248}) into Eq. (\ref{232}) leads to the following set of
identities, following from Eq. (\ref{247}):
\ba
\hat\OO_\nu  = \Bar D_\mu\hat
\te^\mu_\nu+\ha\hat\si^{\rho\si\la}\Bar
R_{\la\nu\rho\si}&+&\fr{1}{2\ka}
\lll  \Bar D_\mu\hat l^{\rho\si}\Bar R^\mu_{~\rho\si\nu}- \hat
l^{\rho\si} \Bar D_\nu \Bar R_{\rho\si}-\ha \Bar D_\rho\hat
l^{\rho\si} \Bar R_{\si\nu}\rrr = 0,\nonumber\\
\label {250}\\
\hat\OO^\mu_\nu = \hat\te^\mu_\nu+\Bar D_\la\hat\si^{\la\mu}_
{~~\nu}- \fr{1}{\ka} \hat l^{\mu\rho} \Bar R_{\rho\nu}&=&  0,	
\label{251}
\\
\hat \OO^{\rho\si}_\nu=\hat\si^{(\rho\si)}_{~~~~\nu}+\Bar D_\mu
\hat\bt_\nu^{~\mu\rho\si}&=&0.
\label{252}
\ea
Equation Eq. (\ref{250}) shows that $\hat\te^\mu_\nu$, the
energy-momentum tensor with respect to a curved background, is in general
not ``conserved"; it is not divergenceless. It is, however, divergenceless if the
background is flat. Equation Eq. (\ref{251}) shows that on a Ricci-flat
background,
$\hat\te^\mu_\nu$  is itself the divergence of a tensor; i.e. it derives from a
superpotential. The generalized Belinfante-Rosenfeld identities may be useful
to check
$\hat\te^\mu_\nu$  and $\hat\si^{\mu\rho\si}$ calculated independently.
Equations Eq. (\ref{250})-(\ref{252}) are a covariant formulation of
Goldberg's \cite{28} identities extended to curved backgrounds.

%________________F. LINEARIZED STRONG ETC.....______

 \vs
\centerline{	\bf{F. Linearized  \clll on a curved background}}
 \vs

In the linear approximation,   we write   $\gmn=
\Bar\gmn + h_{\mu\nu}$ and we omit the overbar on $ \Bar\gmn$; $\Bar
D_\mu$  becomes
$D_\mu$, and terms quadratic in $h_{\mu\nu}$  and $\Bar D_\la
h_{\mu\nu}$  or $ D_\la h_{\mu\nu}$  are neglected. The right-hand side
of Eq. (\ref{235}) becomes
\be 
	\hat l^{\mu\nu}  =\sqrt{-g}(-h^{\mu\nu}+\ha g^{\mu\nu}h)~~~  ,  ~~~
h=g^{\mu\nu} h_{\mu\nu} ;
\label{253}
\ee
indices are now displaced with $\gmn$; for instance, 
$h^{\mu\nu}=g^{\mu\rho}g^{\nu\si}h_{\rho\si}$.  

The right-hand side
of Eq. (\ref{240}), with the superpotential $\hat J^{\mu\nu}$  given by
Eq. (\ref{238}), can now be written entirely in terms of $\hat l^{\mu\nu} $
because, in the linear approximation,  $\De^\mu_{\rho\si}$  defined in
Eq. (\ref{22}), becomes
\be
		\De^\mu_{\rho\si} = \ha\lll   D_\rho h^\mu_\si+ D_\si
h^\mu_\rho-D^ \mu h_{\rho\si} \rrr.
\label{254}
\ee
If we substitute this expression for  $\De^\mu_{\rho\si} $  into Eq.
(\ref{238}), we obtain after a few rearrangements, the perturbed
superpotential density
$ \hat J^{\mu\nu} $ which is linear in $\hat  l^{\mu\nu} $  and its first
derivatives:
\be
		\ka \hat J^{\mu\nu} = \hat  l^{\rho[\mu} D_\rho\xi^{\nu]}  +  
\xi^{[\mu } D_\rho \hat l^{\nu]\rho}- D^{[\mu}\hat l^{\nu]}_\rho \xi^\rho
.	
\label{255}
\ee
	
The left-hand side of Eq. (\ref{240}) contains two terms quadratic in the
perturbations: $t^\mu_\nu$ [cf. Eq. (\ref{221})] and $\si^{\mu[\rho\si]}$ 
[ cf. Eq. (\ref{222})]. These two terms are now neglected. With Eq.
(\ref{254}), the linearized expression for $\hat \ze^\mu$  [cf. Eq.
(\ref{236})] reduces to
\be
4\ka\hat \ze^\mu =Z^{\rho\si}(2D_
\rho\hat l^\mu_\si - D^\mu\hat l_{\rho\si}) - 
\hat l^{\rho\si}(2D_\rho Z^\mu_\si - D^\mu Z_{\rho\si}) +(\hat
l^{\mu\rho} D_\rho Z - D_\rho\hat l^{\mu\rho}  Z).
\label{256}
\ee
Like $\hat J^{\mu\nu}$, $\hat \ze^\mu$  is linear in $\hat l^{\mu\nu}$ 
and its first derivatives. The linearized form of the  conservation law
  (\ref{240}) is thus as follows:
\be
	\int_\Si \lll  \de\hat T^\mu_\nu\xi^\nu+\fr{1}{2\ka}\hat
l^{\rho\si}R_{\rho\si} \xi^\mu+\hat \ze^\mu 
\rrr d\Si_\mu=\int_{\di\Si}\hat J^{\mu\nu}d\Si_{\mu\nu},
\label{257}
\ee
with $\hat J^{\mu\nu}$  given by Eq. (\ref{255}) and $\hat \ze^\mu $  by
Eq. (\ref{256}). The linearized integral identities can also be written in terms
of
$ \de  T^\mu_\nu$  rather than
$ \de\hat T^\mu_\nu$ . Since from Eq. (\ref{253}) we deduce that
\be
	       \de\sqrt{-g}  = \ha \hat l =  \ha \hat h~~~{ \rm  with} ~~~ 
\hat l=\gmn\hat  l^{\mu\nu}  ,~~~ \hat h=\gmn\hat  h^{\mu\nu}, 
\label{258}
\ee
we can replace $\de\hat  T^\mu_\nu $ in Eq. (\ref{257}) by
\be
 \de\hat T^\mu_\nu =\sqrt{-g} \de  T^\mu_\nu  +  \ha
 T^\mu_\nu \hat h~~~, ~~~    \de T ^\mu_\nu=T ^\mu_\nu  -\Bar T
^\mu_\nu ,	
\label{259}
\ee
and obtain, using Einstein's equations for the background,
\be
	\int_\Si\LLL    \de  T^\mu_\nu\hat \xi^\nu+\fr{1}{2\ka}\lll 
R^\mu_\nu\de^\si_\rho-R^\si_\rho\de^\mu_\nu  
\rrr h^\rho_\si\hat \xi^\nu+\hat \ze^\mu
\RRR  d \Si_\mu =\int_{\di\Si}\hat J^{\mu\nu}d \Si_{\mu\nu}  .	
\label{260}
\ee
Equations  (\ref{257}) and   (\ref{260}) are useful forms of the
linearized  integral conservation laws. 

	Simplifications occur when $\ze^\mu$  simplifies; in particular, 
if the background admits con-formal Killing vectors, like in
Robertson-Walker spacetimes, [see Eq. (\ref{242})], in which case Eq.
(\ref{256}) becomes
\be
	8\ka \hat \ze^\mu =(\hat l^{\mu\rho} +\ha g^{\mu \rho }\hat l)\di_\rho
Z - Z D_\rho(\hat l^{\mu\rho}+\ha g^{\mu \rho} \hat l)~~~~~~~{\rm
(\xi^\mu ~ conformal)}.	
\label{261}
\ee
 If $\xi^\mu$  is a homothetic Killing vector, [see Eq. (\ref{244})], then
Eq. (\ref{261}) reduces to
\be
	8\ka \hat \ze^\mu  = -CD_\rho(\hat l^{\mu\rho}+\ha g^{\mu \rho}
\hat l),~~~~~ C={\rm const}    ~~~~~   {\rm (\xi^\mu~ homothetic)}.	
\label{262}
\ee
For Killing vectors of the background, $\ze^\mu =0$. If, in addition, Killing
vectors are tangent to $\Si$, $\bar\xi^\mu d\Si_\mu=0$, as may be the case
in Robertson-Walker spacetimes, Eq. (\ref{260}) reduces then to
\be
	 \int_\Si\lll \sqrt{-g} \de T^\mu_\nu+\fr{1}{2\ka} R^\mu_\nu \hat h  
\rrr\Bar\xi^\nu
d\Si_\mu =\int_{\di\Si}\hat J^{\mu\nu}d \Si_{\mu\nu},
\label{263}
\ee
with $J^{\mu\nu}$  given by Eq. (\ref{255}).	
%___________________III. NOETHER CONSERVATION LAWS__________
\sect {N\oe ther conservation laws}
 
We now return to Eq. (\ref{232}) and consider what happens when arbitrary
$\xi^\mu$'s are replaced by Killing vectors $\Bar\xi^\mu$  of the
background.

%----------------------A. CONSERVED CURRENT $J^\mu$------------------------

\vs
\centerline{	\bf{A.  Conserved current $ J^\mu$}}
\vs 

	$\hat J^\mu$, which contains the physics of the conservation laws, is not, in
general, a conserved vector density since 
\be
		\di_\mu \hat J^\mu  = - \di_\mu \hat\ze^\mu.		
	\label{31}
\ee
 However, when $\xi^\mu$  is a Killing vector
$\Bar\xi^\mu$  of the background, then $Z_{\rho\si}=0$ [cf. Eq.
(\ref{214})] ,
$\hat\ze^\mu=0$, and $\hat J^\mu(\Bar\xi )$ {\it is} conserved.
Hence we can speak about ``physical conservation laws." We should bear in
mind, however, that in general the conserved quantities will depend on the
choice of the background.

$\hat J^\mu $  has been derived in the same way as ``Noether's theorem"
in classical field theory \cite{18}. Thus, by replacing $\xi^\mu$  in strongly
conserved currents by Killing vectors $\Bar\xi^\mu$  of the background, we
obtain Noether conserved vector densities. These are {\it exact} with
mappings on curved backgrounds:
\be
	\hat J^\mu(\Bar\xi)  = \hat \te^\mu_\nu \Bar\xi^\nu  +\hat
\si^{\mu[\rho\si]}
\di_{[\rho} \Bar \xi _{\si]}~~~   , ~~~~     \di_\mu\hat J^\mu(\Bar\xi)  = 0
.	
\label{32}
\ee
	The interpretation of $\hat \te^\mu_\nu$  and $\hat\si^{\mu[\rho\si]}$ 
is suggested by electromagnetic conserved currents in special relativity. For
an electromagnetic field, with 
\be
		\hat\LL _{EM}  = -\fr{1}{16\pi}  \sqrt{-g}F^{\mu\nu} F_{\mu\nu}  ~~~, ~
~~~  F_{\mu\nu} =\di_\mu A_\nu -\di_\nu A_\mu,
\label{33}
\ee
 one finds 
\be
		\hat J^\mu_{EM} = \hat \te^\mu_{\nu EM}\Bar\xi^\nu -\fr{1}{4\pi}
\hat F^{\mu\rho} A^\si \di_{[\rho} \Bar \xi_{\si]} ,
\label{34}
\ee
where $ \Bar \xi^{\nu}$  are Killing vectors of Minkowski space which is
here described in {\it arbitrary coordinates}. In Eq. (\ref{34}), the expression
\be
		\hat \te^\mu_{\nu EM}=
\fr{\di\hat\LL_{EM}}{\di(\di_\mu A_\rho)} \di_\nu
A_\rho-\hat\LL_{EM}\de^\mu_\nu
\label{35}
\ee
represents Pauli's canonical energy-momentum tensor density. It is not the
standard symmetric electromagnetic energy-momentum tensor density 
\be
		\hat T^\mu_{\nu EM}  =  \fr{1}{4\pi}\sqrt{-g}\lll F^{\mu\rho}
F_{\rho\nu} +  \fr{1}{4}\de^\mu_\nu F^{\rho\si} F_{\rho\si}\rrr.	
\label{36}
\ee
 Indeed,
\be
		\hat J^\mu_{EM}  =\hat T^\mu_{\nu EM} \Bar \xi^{\nu} -
\di_\rho\lll  \fr{1}{4\pi}
\hat F^{\mu\rho} A_\nu  \Bar\xi^{\nu}\rrr.	
\label{37} 
\ee
The second term in (\ref{37}) is gauge dependent and its divergence is zero.
It is generally assumed that the appropriate boundary values ensure that this
second term does not contribute to the global conserved quantity. However,
if, with each displacement vector $\Bar \xi^{\nu} $, we associate a gauge
transformation $A_\mu\ra  A_\mu  + \di_\mu \ze$ such that $(A_\mu  +
\di_\mu \ze) \Bar\xi^\mu  = 0$, the gauge dependent term in Eq. (\ref{37})
will disappear.

	The first term,  $\hat T^\mu_{\nu EM} \Bar \xi^{\nu}$, has a proper local
meaning. On a spacelike hypersurface extending to infinity,
\be
\int_\Si\hat J^\mu_{  EM}d\Si_\mu=\int_\Si\hat \te^\mu_{\nu EM} \Bar \xi^{\nu}d\Si_\mu+\int_{\di\Si\ra\infty}\lll  -\fr{1}{4\pi}
\hat F^{\mu\rho} A_\nu  \Bar\xi^{\nu} \rrr d\Si_{\mu\rho}=\int_\Si\hat
T^\mu_{\nu EM} \Bar \xi^\nu d\Si_\mu.
\label{38}
\ee
Here Eq. (\ref{38}) represents the {\it total} energy-momentum for Killing
vectors
$ \Bar \xi^\mu$'s of translations. It gives the total \aaa if 
$ \Bar \xi^\mu$'s describe spatial rotations; the integral of the second term
on the right-hand side of Eq.\,(\ref{32}) represents, in this case, the spin of
the electromagnetic field. This means that $\hat
T^\mu_{\nu EM} \Bar \xi^\nu$  contains also the spin density.
	
By analogy with electromagnetism, we shall give similar interpretations to
the two terms on the right-hand side of Eq. (\ref{32}). $\hat \te^\mu_\nu$ 
is the (relative) energy-momentum tensor density with respect to a given
background for a given mapping and, similarly, $\hat \si^{\mu[\rho\si]}$ 
can be interpreted as the (relative) spin tensor density. As in
electromagnetism, the conserved vector density
$\hat J^\mu$  may not have a well defined local meaning even for a given
mapping. However, $\hat J^\mu$ generates global conservation laws which
are advantageously associated with a superpotential. Global quantities with
appropriate mappings near the {\it boundary} of the domain of integration,
may, and indeed have interesting physical meaning in certain cases as we
shall see below.

%----------------------B. CONSERVATION LAWS IN ETC....------------------------

\vs
\centerline{	\bf{B. Conservation laws in asymptotically flat spacetimes }}
 
 \vs
Locally conserved quantities are related to boundary values through the
superpotential to which we now turn our attention. Global conservation laws
derived from $\hat J^{[\mu\nu]}$ have been discussed in \cite{19}  and in
\cite{21}; they will not be analyzed here. The results of those applications
are, however, illuminating and worth summarizing. They strengthen the
interpretation of $\hat J^\mu$  as a N\oe ther conserved vector density of
energy, linear and angular momentum.
	
Spacetimes that are asymptotically flat admit asymptotic Killing vectors.
Each space  may be mapped on a flat background that is identified with the
spaces themselves at infinity.
	To calculate globally conserved quantities, the mapping can be defined
only asymptotically. To each  Killing vector $\Bar\xi^\mu$  of the
background, the total amount of the corresponding conserved quantity ``in
the whole space at a given time" is the integral of $\hat J^\mu$  over a
spacelike hypersurface $\Si$ extending to infinity:
\be
		P(\Bar\xi^\mu )=\int_\Si \hat J^\mu d\Si_\mu =\int_{\di\Si \ra
\infty}\hat J^{[\mu\nu]}d\Si_{\mu\nu} .	
\label {39}
\ee

%----------------------1.{\it RESULTS AT SPATIAL INFINITY}.------------------------

\vs
\centerline{ {  \tbf {\tsl{  1. Results at spatial infinity}}} }
 \vs
	We may use the asymptotic solution representing an isolated system, as
given in \cite{30}, to calculate energy, linear and \aaaa at
$t=$const. The corresponding quantities $P(\Bar\xi)$  show which
parameters in the  
asymptotic solutions   are commonly interpreted as energy and linear and
\aaaa of such a system. It is worth noting that $\hat
J^{[\mu\nu]}$  provides both linear and \aaaa as does the
pseudotensor of Landau and Lifshitz \cite{26}. Our $\hat J^{[\mu\nu]}$  is,
however, derived from a real N\oe ther conserved vector. In contrast to
the Landau-Lifshitz pseudotensor, it can be calculated in arbitrary
coordinates whereas the Landau-Lifshitz pseudotensor (or the Einstein
pseudotensor for energy and linear momentum) give meaningful results
only in coordinates which become Lorentzian at infinity in such a manner
that
$\gmn \ra \eta _{\mu\nu} + O(r^{-1} )$ (see, however, \cite{15}).

%----------------------1.{\it RESULTS AT NUL INFINITY}.------------------------
\vs
\centerline{ {  \tbf {\tsl{  2. Results at null infinity}}} }
 \vs
	Here, for axisymmetric \cite{14} or general \cite{31} outgoing radiation
asymptotic solutions, it is advantageous to use Newmann-Unti \cite{32}
coordinates $x^\la\equiv (x^0 =t-r\equiv u, r, x^2 , x^3 )$ conformally flat
in
$ x^2 , x^3 $. The solutions have asymptotic symmetries represented by the
Bondi-Metzner-Sachs group \cite{33}. The BMS group contains
supertranslations
$u \ra u+\al (x^2 ,x^3 )$. For the Killing vectors of translations in the
background, we identify $P(\Bar\xi )$, respectively, with the
Bondi \cite{34} mass
$P_0(\Bar\xi )$  and with Sachs \cite{33} linear momentum \cite{35}
$P_k(\Bar\xi )$.
$P_\al(\Bar\xi ) ~ (\al,\bt,....=0,1,2,3)$  behaves like a vector under Lorentz
transformations of coordinates in the flat background, and the fluxes $dP_\al
/du$ are invariant under supertranslations. Similarly, for Killing vectors of
spatial rotations in the background, $P(\Bar\xi )$ is the same \cite{21} as
the  standard definition of the \aaa $L_k(\Bar\xi)$ \cite{22}, 
without an ``anomalous factor 2". The angular momentum transforms as a
vector for rotations in the background but $dL_k /du$ depends on the
mapping and, in particular, on supertranslations.
	
The conserved quantities $P(\Bar\xi )$ have one outstanding property worth
noting. They are given by a superpotential, not an ``asymptotic
superpotential". That is, $P(\Bar\xi )$ is obtained from a differential
conservation law and is directly related through Einstein's equations to the
energy-momentum tensor of the matter. No other differential conservation
law has been given so far (with or without a background) that gives the
standard expressions of the total energy, linear and \aaa at
null infinity.

%----------------------C  Linearized conservation laws}.------------------------

\vs
\centerline{{\bf  C. Linearized conservation laws }}
 \vs
	In the linear approximation, the formulas of Section II E are valid.
N\oe ther's conserved currents follow from Eq. (\ref{257}) or  
(\ref{260}) by replacing
$\xi^\mu$ with Killing vectors $\Bar\xi^\mu$  for which $\ze^\mu=0$.
Thus, the linearized form of the global N\oe ther conservation laws
  (\ref{39}) becomes
\be
	\de P(\Bar\xi ) = \int_\Si \sqrt{-g}\LLL      \de  T^\mu_\nu +\fr{1}{2\ka}\lll 
R^\mu_\nu\de^\si_\rho-R^\si_\rho\de^\mu_\nu  
\rrr  h^\rho_\si  
\RRR \Bar\xi^\nu d \Si_\mu
=\int_{\di\Si}\hat   J^{\mu\nu}(\Bar\xi )d \Si_{\mu\nu}  .	
\label{310}
\ee
 with
\be
		 	\ka  \hat J^{\mu\nu} =  \hat  l^{\rho[\mu} D_\rho\Bar\xi^{\nu]}  +  
\Bar\xi^{[\mu } D_\rho  \hat  l^{\nu]\rho}- D^{[\mu} \hat  l^{\nu]}_\rho
\Bar\xi^\rho .	
\label{311}
 \ee
 %___________________III. STRONGLY CONSERVED CURRENTS__________

\sect  {Conservation laws in cosmology with  respect to\\
\centerline { de Sitter backgrounds$~~~~~~~~~~~~~~~~~$}}

%----------------------A.  SPATIALLY CONFORMAL ETC....------------------------

\centerline{{\bf  A. Spatially conformal mappings on de Sitter space }}
\vs 

Strongly or weakly perturbed Robertson-Walker spacetimes are related, by
definition, to a Robertson-Walker background. Robertson-Walker spacetimes
admit six Killing vectors, each of these vectors generate a conserved
N\oe ther current \cite{36}.
 
	We may, however, map Robertson-Walker spacetimes (perturbed or not) on
a de Sitter space  which has ten Killing vectors and thus ten conserved
currents. It is interesting to ask what are the four N\oe ther currents for
Killing vectors of  the de Sitter space that are not Killing vectors of a
Robertson-Walker spacetime. They correspond to energy and linear
momentum. To elucidate this, we map closed ($k=1$), flat ($k=0$)  or open 
($k=-1$)
hypersurfaces (at constant cosmic time) of Robertson-Walker spacetimes on
the corresponding hypersurfaces, with the same topology, in de Sitter
spaces. 
	
Let Robertson-Walker spacetimes be described in coordinates $(t,x^k)$ in
which the metric reads
\be
	ds^2 =\gmn dx^\mu dx^\nu=\phi^2 dt^2 +g_{kl }dx^k dx^l =\phi^2 dt^2 
- a^2 f_{kl} dx^k dx^l ,	
\label{41}
\ee
 where $f_{kl}(x^m)$  have particular forms for closed, flat or open
$t$=const  hypersurfaces; $x^k$  may be any of suitable coordinates, $\phi$
and $a$ are functions of $t$. The metric of  the de Sitter background in these
coordinates has a similar form
\be
	\Bar {ds}^2 =\Bar\gmn dx^\mu dx^\nu =\psi^2 dt^2 +\Bar g_{kl} dx^k
dx^l =\psi^2 dt^2  - \Bar a^2 f_{kl} dx^k dx^l;
\label{42}
\ee
here, $\psi$ and $\Bar a$  are also functions of $t$. The ``cosmic (proper)
time" $T$ is thus given by $dT=\phi(t)dt$ in the Robertson-Walker
spacetimes and by $dT=\psi(t)dt$ in de Sitter space. Hypersurfaces with the
same $t$ are mapped on one another. Choosing both functions $\phi$ and
$\psi$ fixes the mapping of the cosmic times up to a constant. For the
moment we shall fix neither of them.

%----------------------B.  KILLING VECTORS ETC....------------------------

\vs
\centerline{{\bf  B. Killing vectors of the de Sitter background }}
 
\vs
The ten Killing vectors of the de Sitter background, $\Bar\xi ^\mu =
(\Bar\xi ^0 ,\Bar\xi ^k )$, satisfy the Killing equations
\be
		\Bar Z_{\mu\nu}  = \Bar D_\mu  \Bar\xi _\nu +\Bar D_\nu  \Bar\xi
_\mu =0,	
\label{43}
\ee    
which in the 1+3 decomposition given by Eq. (\ref{42}) imply
\ba
	Z_{00} =0   ~\Ra~   \Bar\xi^0&=&\fr{1}{\psi}\td\xi^0(x^k),	
\label{44}\\
	Z_{0k} =0  ~\Ra ~\dot{\Bar\xi^k}     &=& -\psi^2\Bar
g^{kl}\na_l\Bar\xi^0,
\label{45}
\ea
where $\td\xi^0 $  is a function whose equation is given below [see
Eq. (\ref{49})], $\na_l$ is a $\Bar g_{kl} $ (or $g_{kl}$, or $f_{kl}$) 
covariant derivative, and a dot denotes a derivative with respect to $t$. It
may be useful to remind the reader that indices are displaced by $\Bar
g_{\mu\nu}$. Finally, the spatial part of the Killing equations gives
\be
	          -\fr{1}{\Bar a^2}\Bar
Z_{kl}=f_{mk}\na_l\Bar\xi^m+f_{ml}\na_k\Bar\xi^m+2\psi\Bar
Hf_{kl}\Bar\xi^0=0,
\label{46}
\ee
where
\be
			\Bar H = \fr{\dot{\Bar a}}{\psi\Bar a}		
\label{47}
\ee
is the Hubble ``constant" of de Sitter space; $\Bar H$  satisfies the
relation 
\be
			\fr{1}{\psi} \dot{\Bar H} =  \fr{k}{\Bar a^2},		 
\label{48}
\ee
 which follows from EinsteinÕs equations or, as the
integrability condition of  Eq. (\ref{43}). If we take a partial $t$-derivative of  
Eq. (\ref{46}) and make use of   Eq. (\ref{45}), we obtain
\be
		\na_{kl} \Bar\xi^0 + kf_{kl}  \Bar\xi^0  = 0~~~  { \rm  or  }~~~    
\na_{kl}\td\xi^0 + kf_{kl} \td\xi^0  = 0 .
\label{49}
\ee
This equation can be solved. Having $\td\xi^0$, we can obtain $\td\xi^k$ 
 from Eqs. (\ref{45}) and (\ref{46}). Explicit expressions for
$\Bar\xi^\mu$  and finite group transformations are given in Weinberg
\cite{37}. Using Weinberg's coordinates {\it adapted} to $t=$const  slices,
$f_{kl}$ becomes 
\be
f_{kl}  =\de_{kl}  +\fr{kx^kx^l}{1-kr^2} ,~~~  f^{kl} = \de^{kl} -kx^k x^l ,
~~~   r^2 =(x^1)^2 +(x^2)^2 +(x^3)^2.	
\label{410}
\ee
Any $\Bar\xi^\mu$  is a linear combination with constant coefficients of
the following ten vectors: 

(a) Quasitranslations in $t$=const:
\be
	 	  {\Bar\xi}_{(r)} ^0 = 0 ,  ~~~   \Bar\xi^k_{(r)}  
=\de^k_r\sqrt{1-kr^2}   ,~~~ r=1,2,3.
\label{411}
\ee

(b) Quasirotations in $t$=const:
\be
		 \Bar\xi^0_{[rs]} =0, ~~~  \Bar\xi^k_{[rs]}     = \de^{kr} x^s  - \de^{ks}
x^r,~~~ r,s=1,2,3.	
\label{412}
\ee

(c) Time quasitranslations:
\be
			 \Bar\xi^0_{(0)} =\fr{1}{\psi}\sqrt{1-kr^2}, ~~~  \Bar\xi^k_{(0)}  = -\Bar
H x^k
\sqrt{1-kr^2}.	
\label{413}
\ee

(d) Lorentz quasirotations: 
\ba
	 \Bar\xi^0_{[r]} = \fr{1}{\psi}x^r,~~~     k=0  ~~~&\Ra&~~~ 
\Bar\xi^k_{[r]}  =\Bar H 
\LLL  \ha \de^{kr}(r^2-\tau^2)-x^k x^r \RRR,
\label{414}\\
		                    k=\pm 1~~~&\Ra&~~~  \Bar\xi^k_{[r]}= \Bar H \LLL
k\de^{kr} - x^k x^r \RRR,
\label{415}
\ea
where in Eq. (\ref{414}),
\be
			\tau = \fr{\psi}{\dot{\Bar a}}~~~~~~(k=0).
\label{416}
\ee
 The Killing vectors   (\ref{411}) and   (\ref{412}) are also the Killing
vectors of Robertson-Walker spacetimes. The vectors
 (\ref{413}),  (\ref{414}), and  (\ref{415}) are conformal Killing
vectors of Robertson-Walker spacetimes.

%---------------------C.  SUPERPOTENTIALS AND ETC....------------------------
\vs

\centerline{{\bf  C. Superpotentials and conserved vectors }}
 \vs
 
	To obtain the superpotentials we follow the calculations outlined in Section
II C. With the metric components $\gmn$  of Eq. (\ref{41}) and $\Bar\gmn$ 
of Eq. (\ref{42}), we calculate the quantities $ l^{\rho\si} =\hat  l^{\rho\si}
/\sqrt{-g} $,
\be
		l^{00} = \fr{1}{\phi^2}\lll  1-\fr{\phi\Bar a^3}{\psi a^3} \rrr~~~,   ~~~ 
l^{kl} = g^{kl}\lll  1-\fr{\psi\Bar a }{\phi a } \rrr,
\label{417}
\ee
and the Christoffel symbols $\Ga^\la_{\mu\nu}$  and
$\Bar\Ga^\la_{\mu\nu}$  and their  differences $\De^\la_{\mu\nu}$,
\be	
 \Ga^0_{00}   =\fr{\dot\phi}{\phi}  ,~~~       \Bar \Ga^0_{00} = 
\fr{\dot\psi}{\psi} ~~~\Ra~~~    \De^0_{00} = \fr{\dot\phi}{\phi} -
\fr{\dot\psi}{\psi}\equiv \phi\TT.	
\label{418}
\ee
The function $\TT$ just defined describes a relative shift in times measured
in Robertson-Walker cosmic time units. Next,
\be
\Ga^k_{0l} =\phi H\de^k_l       , ~~~    \Bar\Ga^k_{0l} = \psi \Bar H\de^k_l
~~~\Ra~~~     \De^k_{0l}  = \phi (H- \fr{\psi}{\phi}\Bar H)\de^k_l=\phi \HH\de^k_l,
\label{419}
\ee
where
\be	
			H =\fr{\dot a}{\phi a};
\label{420}
\ee
$\HH$ is the relative Hubble function measured in Robertson-Walker cosmic
time. Finally, 
\be
\Ga^0_{kl }= -\fr{H}{\phi}   g_{kl} ,~~~   \Bar\Ga^0_{kl }=
-\fr{\Bar H}{\psi}   \Bar g_{kl}~~~\Ra~~~\De^0_{kl } = -\fr{1}{\phi}\lll  
H-\fr{\Bar a^2\phi}{a^2\psi}\Bar H \rrr g_{kl}.
\label{421}
\ee
With $\Bar\xi^\mu$  given by Eqs.\,(\ref{411})-(\ref{416}), $l^{\mu\nu}$ 
by Eq. (\ref{417}), $\De^\la_{\mu\nu}$  by Eqs.\,(\ref{418})-(\ref{421}), the 
superpotential, defined in Eq. (\ref{238}), has the  form:
\ba
2\ka J^{ 0k} & =& \AA\Bar g^{kl} \na_l \Bar\xi^0 + \BB \Bar\xi^k ,	
\label{422}\\
 2\ka J^{ kl} & =& \CC\Bar g^{m[k} \na_m\Bar\xi^{l]},  	
\label{423}
\ea
where $\AA,\BB$ and $\CC$ are functions of $t$: 
\ba
\AA(t)&=& -\fr{\Bar a^2}{a^2}  +  2\fr{\psi\Bar a^3}{\phi a^3}-\fr{\psi^2
}{\phi^2 },\nonumber\\
\BB(t) &=& \lll  3\fr{\Bar a^2}{a^2}  - 2\fr{\psi\Bar a^3}{\phi
a^3}-\fr{\psi^2 }{\phi^2 } \rrr \fr{\Bar H}{\psi}-\fr{4}{\phi}\HH  ,  \\
			       C (t)&=&  2\fr{\Bar a^2}{a^2} - 2\fr{\psi\Bar a^3}{\phi
a^3} .\nonumber		
\label{424}
\ea
The components of the conserved vector density $ \hat J^\mu$  can be
calculated either from the superpotential since $ \hat J^\mu = \di_\nu
 \hat J^{\mu\nu}$  or directly from Eq. (\ref{233}). With the usual
notations for the
$T^\mu_\nu$ components
\be
	T^0_0   = \rho,~~~     T^k_l = - \de^k_l P,  ~~~{\rm and}~~~  
\Bar T^\mu_\nu  =
\La\de^\mu_\nu,		
\label{425}
\ee
the zero component of the current then reads
\be
  	J^0  =  \LLL  \lll  \rho-\fr{\psi\Bar a^3}{\phi
a^3}\fr{\La}{\ka} \rrr-\fr{1}{2\ka}l\La-\fr{3}{\ka}\HH^2 \RRR
\Bar\xi^0\equiv \UU(t)\Bar\xi^0,
\label{426}
\ee
where
\be
		l=l^{\rho\si} \Bar g_{\rho\si}  =  3\fr{\Bar a^2}{a^2}  - 4 \fr{\psi\Bar
a^3}{\phi a^3} + \fr{\psi^2
}{\phi^2 } = \CC-\AA,	
\label{427}
\ee
and the spatial part is given by
\ba
J^k &= & \LLL  \lll  -P-\fr{\psi\Bar a^3}{\phi
a^3}\fr{\La}{\ka}
\rrr-\fr{1}{2\ka}l\La+\fr{3}{\ka}\HH^2-\fr{3\phi}
{2\ka\psi}\lll \fr{\Bar a^2}{a^2} -\fr{\psi^2}{\phi^2}\rrr(\TT+\HH)\Bar
H\RRR\Bar\xi^k\nonumber
\\ 
&+&\ha\fr{  \phi}{ \ka\psi}\lll \fr{\Bar
a^2}{a^2} -\fr{\psi^2}{\phi^2}\rrr(\TT+\HH)\Bar g^{kl}\na_l(\psi
\Bar\xi^0).
\label{428}
\ea
The first parentheses in the brackets of Eqs. (\ref{426}) and   (\ref{428}) 
represents
the ``relative mass-energy density" and ``relative pressure" respectively.
The second term is the coupling to the background. The other terms are
associated with field energy and helicity and they depend on the mapping
of the time axes.

%---------------------D.  MAPPINGS------------------------

\vs
\centerline{{\bf  D. Mappings }}
\vs 

	As a consequence of Eq. (\ref{426}) and $ \Bar\xi^0$'s as given in
Eqs. (\ref{411})-(\ref{419}), the conserved quantities in a volume $V$
enclosed by a sphere of radius $r$ are all equal to zero except the ``energy" 
$P_0$, associated with time quasitranslations $ \Bar\xi^0_{(0)}$  given by
Eq. (\ref{413}). The ``energy" reads
 \be
		P_0  = \fr{4\pi}{3} \Bar a^3r^3\UU(t)\fr{\phi}{\psi},
\label{429}
\ee
where $\UU$ is given by Eq. (\ref{426}). The most appealing mapping is one
that gives $\UU=0$ so that $P_0=0$. With such a mapping there are ten
conserved quantities for {\it perturbations} of Robertson-Walker spacetimes
only; it adds ``energy-momentum" to the {\it perturbed} Robertson-Walker
spacetimes that have no quasitranslation invariance. 

	One may also consider a conformal mapping, for which we take
\be
	\psi=1, ~~~      \phi=\fr{a}{\Bar a}  ,	
\label{430}
\ee
 so that 
\be
		ds^2  =  \fr{a^2}{\Bar a^2}  \Bar {ds}^2 .	
\label{431}
\ee
Then,
\be
	\UU(t)
=(\rho-\phi^{-4}\fr{\La}{\ka})+2\phi^{-4}(1-\phi^2)\fr{\La}{\ka}-
\fr{3}{\ka}\HH^2.
\label{432}
\ee
In this case the total energy of a closed space $(k=1)$ is also zero, but for
open or flat sections it is infinite. The mean Òenergy densityÓ 
$P_0/[(4\pi/3)\Bar a^3r^3]$  of a $k=0$ Robertson-Walker spacetime
mapped on de Sitter is given by $\UU(t)\sqrt(-g)/\sqrt(-\Bar g)$. The mean
``energy density" in a $k=-1$ space is, however, infinite because $P_0$ 
grows faster than the proper volume as $r\ra\infty$.

%___________________II. TRASCHEN INT.CONST......__________

\sect {Traschen's integral constraints}

%---------------------A. EQUATIONS FOR INTEGRAL CONSTRAINTS. -----------

\centerline{{\bf  A. Equations for integral constraint vectors }}
\vs
Let us now go back to Traschen's integral constraints that were written
in Eq.\,(\ref{13}) for spatially localized linear perturbations. For general
linear perturbations, Traschen \cite{5} showed that for certain vectors
$V^\mu$, called ``integral constraint vectors" (ICV's), that satisfy 12
equations on a particular hypersurface $S$, there exist Gauss-like integrals of
the form
\be
	\int_\Si\de T^\mu_\nu \hat V^\nu d
\Si_\mu=\int_{\di\Si}\hat B^{\mu\nu}d  \Si_{\mu\nu},
\label{51}
\ee
where $B^{\mu\nu}$  is given in terms of $h_{\mu\nu}$, $V^\mu$,  and
their first order derivatives. If the perturbed metric gives no contribution to
the right-hand side of Eq.\,(\ref{51}), the expression reduces to Traschen's
integral constraints   (\ref{13}). At first sight,  Eq.\,(\ref{51}) appears to
be a  conservation law for linear perturbations similar to
Eq.\,   (\ref{260}), some terms of the left-hand side of Eq.\,(\ref{260}) having
been transformed into boundary integrals.
	
The 12 Traschen equations for ICV's were deduced from Einstein's
constraint equations \cite{38}. We shall here show that Traschen's ICV 
equations can be derived from the  \clll   (\ref{260}).
	
The problem is to find the conditions on $\xi^\mu$'s for which 
Eq. (\ref{260}) takes the form  (\ref{51}). In doing so, we shall not only
obtain the Traschen equations for $V^\mu$,  but also find under what
conditions Eq. (\ref{51}) holds on a {\it family} of hypersurfaces $S$ rather
than on a particular hypersurface.

	Let us write  Eq. (\ref{260}) in synchronous coordinates around
$S$.  In these coordinates, $S$  is defined by $t$=const  and the metric
takes the form
\be
	ds^2  = dt^2  + g_{kl}(t,x^m) dx^k dx^l,~~~~    k,l,m,...=1,2,3.	
\label{52}
\ee

	It is always possible to keep the gauge synchronous for the perturbations,
namely, to take
\be
		h_{00}  = h_{0k}  = 0,		
\label{53}
\ee
because $h_{00}$  and $h_{0k}$  depend on the mapping above and below
$S$. Here we are interested in conditions on one particular hypersurface
$S$ (to begin with). On $t$=const,   Eq. (\ref{260}) can be written as
\be
	\int_\Si\LLL  \de T^0_\nu\xi^\nu+\2k \lll  R^0_\nu\xi^\nu h
-R^\si_\rho h^\rho_\si\xi^0\rrr +\ze^0\RRR dV=\int_{\di\Si}J^{0k}dS_k,
\label{54}
\ee
where
\be
   dV=\sqg d \Si_0 =\sqrt{-g}dx^1 dx^2 dx^3~~~{   \rm   and}~~~      dS_k  =
\sqg d  \Si_{0k} =\sqrt{-g} \ep _{klm} dx^{[l}dx^{m]} .
\label{55}  
\ee
The component $\ze^0$  [cf. Eq. (\ref{256})] is linear in $Z_{\mu\nu}$  and
$D_\rho Z_{\mu\nu}$. It is also linear in $h_{mn} , ~\dot h_{mn}  =\di_t
h_{mn}$  and $\na_kh_{mn}$ (the covariant derivatives of $h_{mn}$ with
respect to the three-metric $g_{kl}$). Thus, $\ze^0$ is of the form
\be
  \ze^0   = A^{kl} h_{kl} +B^{mkl }\na_m h_{kl} +C^{kl} \dot h_{kl}
=\na_m(B^{mkl }  h_{kl}) +E^{kl} h_{kl} +C^{kl} \dot h_{kl}.
\label{56}
\ee	
Inserting Eq. (\ref{56}) into Eq. (\ref{54}) and taking account of Eq.
(\ref{53}), we obtain an expression of the form
\be
 \int_\Si\LLL  \de T^0_\nu\xi^\nu+\2k Y^{kl}\td h_{kl} +\4k
Z^{kl}\di_t \td h_{kl}\RRR dV=\int_{\di\Si}\lll J^{0m}-B^{mkl}h_{kl}   \rrr
dS_m,
\label{57}
\ee
in which
\be
		\td h_{kl}  = h_{kl} -g_{kl} h^m_m   ,		
\label{58}
\ee
and the $Z^{kl}$'s are the spatial components of the $Z_{\mu\nu}$  tensor
defined in Eq. (\ref{214}). The left-hand side of Eq. (\ref{57}) takes the form
 (\ref{51}) when the factors of $\td h_{kl} $  and of its time derivative
$\di_t \td h_{kl}$  vanish: 
\be
	Z_{kl } = 0, ~~~   Y_{kl}  = 0 .		
\label{59}
\ee
The first of these equations can be written in a 1+3 decomposition as
\be
		Z_{kl} = \na_k \xi_l +\na_l \xi_k +\dot g_{kl}\xi^0  = 0		
\label{510}
\ee
(remember that in general $D_k \neq \na_k$). With $Z_{kl} =0$, the
equation $Y_{kl} =0$  reduces to
\be	
Y_{kl} =  \ha \lll  \na_kZ^0_l+  \na_l Z^0_k\rrr+\fr{1}{4}\dot
g_{kl}Z^0_0-\lll R_{kl}+g_{kl} G^0_0   \rrr\xi^0-\ha R^0_m\xi^mg_{kl}=0,
\label{511}
\ee
where $G^0_0$  is a component of Einstein's tensor. Accordingly,  if  ICV's
satisfy  Eqs. (\ref{510}) and   (\ref{511}), then Eq. (\ref{54}) or  
(\ref{57}) has the form   (\ref{51}). It is now easy to see that  Eqs.
(\ref{510}) and   (\ref{511}) are equivalent to Traschen's equations (3a)
and (3b)
\cite{39}. Inserting the explicit expressions of $B^{mkl}$  into Eq. (\ref{57}),
we obtain
\be
\int _\Si \de T^0_\nu\xi^\nu dV=\int_{\di\Si}\lll
J^{0k}-\4k  h^m_mZ^k_0  \rrr dS_k,
\label{512}
\ee
where $J^{0k}$  is given in terms of $\xi^\mu,~h_{\mu\nu}$,  and their
first order derivatives by Eq. (\ref{255}). 
	
How is Eq. (\ref{512}) modified if we consider perturbations in a
non-synchronous gauge? The answer is: instead of  Eq. (\ref{512}), 
Eq. (\ref{54}) becomes 
\be
	\int_\Si\LLL  \de T^0_\nu\xi^\nu+\2k R^0_k\lll  h^0_0\xi^k 
-2h^k_0\xi^0\rrr \RRR dV=\int_{\di\Si} \LLL  J^{0k} -\4k
h^m_mZ^k_0+\4k\lll  Z^k_0h^0_0-Z^0_0 h^k_0 \rrr\RRR dS_k.
\label{513}
\ee
Equation  (\ref{513}) shows that if Eqs. (\ref{510}) and  (\ref{511}) 
hold {\it and} if $R^0_k =0$ in synchronous coordinates,  Eq. (\ref{260})
has the desired form   (\ref{51}) independently of any gauge condition, as
pointed out by Traschen. Robertson-Walker spacetimes have that property,
but, in general, $R^0_k$ does not vanish. In a synchronous gauge,
Eq. (\ref{260}) has the form of Eq. (\ref{51}) not only on a particular $S$, but
on all nearby hypersurfaces.
\vs

%---------------------B. ICV''s IN RW SPACETIMES-----------

\centerline{{\bf  B. ICV's in Robertson-Walker \ssss }}
 
 \vs
	With a metric of the form   (\ref{41}) and with $\phi=1$,  
Eq. (\ref{510}) can be written as
\be
	-\fr{Z_{kl}}{a^2} =f_{mk}\na_l\xi^m+f_{ml}\na_k\xi^m +2f_{kl}H\xi^0= 0,
~~~{\rm  where}~~~   H = \fr{\dot a}{a} 
\label{514}
\ee
and $Y_{kl} =0$ or, equivalently, 
\be
	Y_{kl}  - \ha a^2\di_l\lll   \fr{Z_{kl}}{a^2}\rrr\equiv \na
_{kl}\xi^0+kf_{kl}\xi^0=0.
\label{515}
\ee
Equations  (\ref{514}) and  (\ref{515}) are Traschen's equations (15a)
and (15b) in \cite{5}. 

 We notice that   Eq. (\ref{515}) for $\xi^0$  is the
same as   Eq. (\ref{49}) for $ \Bar\xi^0$, and that  
Eq. (\ref{514}) for $\xi^0$  and $\xi^k$  is the same as   Eq.
(\ref{46}) for $
\Bar\xi^0$  and $\Bar\xi^k$  in the de Sitter space provided that
\be
\psi=\fr{H}{\Bar H}.
\label{516}
\ee
Therefore, a set of solutions for Traschen's equations is given by the ten
Killing vectors $ \Bar\xi^\nu_{(a)} (a= 1,2,...10)$  of the de Sitter space
  (\ref{411})-(\ref{416}) with $\psi$ replaced by $H/\Bar H$. Linear
combinations of the ten Killing vectors  $ \Bar\xi^\nu_{(a)}$'s, with
coefficients that are functions of $t$, are also solutions of Traschen's
equations. In effect, the ten ICV's, say $V^\mu_{(a)}$ , given by Traschen,
are the following combinations of the Killing vectors:
\be
			V^\mu _{(a)} =\psi\bar\xi^\mu _{(a)}=\fr{H}{\Bar H}\Bar\xi^\mu _{(a)},
\label{517}
\ee
 with the exception of quasi-Lorentz rotations in the flat Robertson-Walker
spacetime $(k=0)$ for which Traschen's ICV's are equal to 
\be
		                 V^\mu _{[r]}  = \psi\lll \Bar \xi^\mu_{[r]}   + \ha \tau^2  \Bar
\xi^\mu _{(r)} \rrr,	
\label{518}
			\ee 
where $\Bar \xi^\mu_{[r]} $   is given by Eq. (\ref{414}),  $\Bar
\xi^\mu_{(r)}
$  by Eq. (\ref{411}), and $\tau$ by Eq. (\ref{416}).

	Equations   (\ref{517}) and (\ref{518}) suggest, and it has been shown
explicitly in \cite{41}, that in fact Traschen's integral constraints  
(\ref{13}) are conservation laws for a perturbed Robertson-Walker spacetime
{\it mapped on a de Sitter spacetime} and with the mapping given by the
conditions	
\be
		\phi=1,~~~\psi=\fr{H}{\Bar H}.
\label{519}
\ee
This is at variance with Traschen and Eardley's \cite{6} interpretation of
Eq. (\ref{13})  as conditions of energy-momentum
conservation with respect to the Robertson-Walker background.
\vs
\ni\Large{{\bf Acknowledgements}}
\vs 
\normalsize

 J.B. acknowledges financial support from the  Royal Society
and the support from the grants Nos. GA\v CR-202/96/0206 and
GAUK-230/1996 of the Czech Republic and the Charles University.
	J.K. acknowledges interesting conversations with N. Deruelle and J.  P. Uzan
about Traschen's integral constraints. He also thanks A. Petrov of The
Sternberg Astronomical Institute in Moscow  for checking many of the
formulas and for useful advise.


\begin{thebibliography}{99}
\bibitem{27} The
Belinfante-Rosenfeld identities generalized to curved backgrounds  are
given in Section II D.
\bibitem{39} Equations (3a) and (3b) are in  [5]. There is a printing
mistake in those equations which has been pointed out by Tod [23].
Incidentally, Tod has shown that Traschen's equations are conditions for
$S$ to be embeddable in a space of {\it constant curvature} of which the
ICV's are the Killing vectors. See in this connection Section V B below.
\end{thebibliography}
\end{document}